\documentclass[twocolumn,
pra,secnumarabic,nobalancelastpage,
amsmath,amssymb,superscriptaddress,nofootinbib]
{revtex4-2}

\usepackage{graphicx} 
\usepackage{xcolor}
\usepackage{hyperref}

\newcommand{\req}[1]{Eq.\,(\ref{#1})} 
\newcommand{\rs}[1]{section~\ref{#1}} 
\newcommand{\ra}[1]{appendix~\ref{#1}} 
\newcommand{\rf}[1]{Fig.\,\ref{#1}}

\begin{document}

\title{Non-Hermitian photon number 
filtering using N00N state Bloch oscillations}

\author{Sean Crowe}
\affiliation{Naval Information Warfare Center Pacific, San Diego, CA, 92152, United States}
\author{Joanna Ptasinski}
\affiliation{Naval Information Warfare Center Pacific, San Diego, CA, 92152, United States}
\author{Melvin Pascoguin}
\affiliation{Naval Information Warfare Center Pacific, San Diego, CA, 92152, United States}
\author{Stefan Evans}
\affiliation{Naval Information Warfare Center Pacific, San Diego, CA, 92152, United States}

\begin{abstract}
We explore Bloch oscillations of $N=1$ and $N=2$ photonic N00N states, using an array of linearly growing effective index waveguides with a simulated asymmetric loss profile. Starting with equal probability input $N=1$ and $N=2$ N00N states, and siphoning off a portion of the waveguides near the half Bloch period, we selectively output specific photon number states. Tuning the input phase, we achieve dynamic switching between   2-photon dominated (80\% of the output) and 1-photon dominated (90\% of the output) cases. This offers a path to improved state preparation in photonic circuits used in computing and networking. 
\end{abstract}


\maketitle

\section{Introduction}

Entangled photonic N00N states are well-known to exhibit an interference pattern analogous to that of a reduced effective de Broglie wavelength~\cite{Jacobson_1995, Walther_2004, Hofmann_2007, Dowling_2008, Afek_2010}, enabling an enhanced phase sensitivity for quantum metrology~\cite{Giovannetti_2011}. This interferometry framework encompasses sub-diffraction limit lithography~\cite{Boto_2000} and imaging~\cite{Ono_2013}, as well as sub shot noise precision in Sagnac interferometers~\cite{Fink_2019, Silvestri_2024}. N00N states have also been shown to produce nontrivial interference effects in periodic structures~\cite{Selim_2025}.

Here we consider a series of waveguides, in which photonic N00N state Bloch oscillations have been shown to exhibit photon-number dependent interference profiles~\cite{otherBO, Bromberg_2010, Lebugle_2015, Hui_2026}. Especially notable is the tendency of $N=1$ N00N states to exhibit interference fringes in which the photon density is localized to one side of the platform depending on the input phase. In contrast, the $N=2$ states are evenly distributed across both sides of the platform. This is due to Bloch oscillations of $N=1$ N00N states behaving as coherent states, and in the $N>1$ case as incoherent beams input into separate waveguides~\cite{Bromberg_2010}.

In this work, we propose and analyze a filtration scheme based on this asymmetry between $N=1$ and $N=2$ N00N state Bloch oscillation photon density profiles. By incorporating loss in half of the waveguide platform at the half Bloch period, we selectively filter  $N=1$ or $N=2$ photon number states. By varying the input phase, we control which side of the platform the 1-photon sector is localized to, and thus whether it is on the transmissive or lossy side. Thus we switch between filtered $N=1$ and $N=2$ photon subspaces distributed over two waveguides, which can also be used for de-multiplexing.

In~\rs{overview}, we summarize some of the relevant properties of N00N state Bloch oscillations in zero loss waveguide arrays. We then implement asymmetric loss on one side of the platform in~\rs{asymmetric}, which we simulate via evolution of a reduced density matrix. Our results are summarized in the conclusion~\rs{concl} with a note on avenues for further optimization.

\section{Bloch oscillations in zero loss platforms}
\label{overview}

We first give an overview of Bloch oscillations in  waveguides with no loss, following the steps in \cite{Bromberg_2010}. We consider the Hamiltonian 
\begin{equation}
\label{Ham}
\hat H=
\!\!\!\!\!
\sum_{\nu=-N/2}^{N/2}
\!\!\!\!\!
\nu B \hat a^{\dagger}_{\nu}\hat a_{\nu}
+C \left(\hat a^{\dagger}_{\nu+1}\hat a_{\nu}+\hat a^{\dagger}_{\nu}\hat a_{\nu+1}\right)
\;,
\end{equation}
where we sum over $N$ waveguides, which must be sufficiently large that surface effects can be neglected. Here $\nu$ is the waveguide number, and the commutator $\left[\hat a_{\nu},\hat a_{\mu}^{\dagger}\right]=\delta_{\mu \nu}$. The resulting equations of motion
\begin{equation}
\label{eqn:HeisenbergEquations}
-i \frac{\partial \hat a^{\dagger}_{\mu}}{\partial z}=
\mu B\hat a^{\dagger}_{\mu}+C\left(\hat a^{\dagger}_{\mu+1}+\hat a^{\dagger}_{\mu-1}\right)
\;,
\end{equation}
where the difference between neighboring propagation constants $
B=\frac{2\pi}{\lambda}\Delta n_{\rm eff}$. The effective index $n_{\rm eff}$ grows linearly across  the waveguides, and the resulting Bloch period 
\begin{equation}
\lambda_B=2\pi/B
\;.
\end{equation}
The mode coupling coefficient
\begin{equation}
C=\pi/\left(2L_t\right)
\;,
\end{equation}
where $L_t$ is the distance over which light transfers between adjacent waveguides.

We consider the initial state input into adjacent waveguides $\mu,\nu$ in the center of the platform
\begin{align}
\label{psiInitial}
|\psi_0\rangle=
\frac{
\alpha\Big(|1_\mu0_{\nu}\rangle\!+\!e^{i\phi}|0_{\mu}1_{\nu}\rangle\Big)
+\beta\Big(|2_{\mu}0_{\nu}\rangle\!+\!e^{2i\phi}|0_{\mu}2_{\nu}\rangle\Big)
}
{{\cal N}}
,
\end{align}
with normalization ${\cal N}=\sqrt{2(|\alpha|^2+|\beta|^2)}$, and the tensor product with vacuum modes for the adjacent waveguides implicit. In terms of creation operators, the evolution in $z$ along the waveguides follows
\begin{align}
\label{aEvolve}
\hat a^\dagger_\mu(z)=
\sum_{\mu^\prime} U_{\mu,\mu^\prime}(z)
\hat a^\dagger_{\mu^\prime}(0)
\;,
\end{align}
where $U_{\mu,\mu^\prime}(z)$ is the Green function used in~\cite{Bromberg_2010}:
\begin{align}
\label{Ugreenform}
U_{\mu,\nu}(z)=
e^{i\frac\pi 2(\mu-\nu)}e^{iBz/2(\mu+\nu)}
J_{\mu-\nu}\Big(4C/B \sin(Bz/2)\Big)
\;,
\end{align}
and $J$ is the Bessel function of the first kind. The resulting photon density
\begin{align}
\label{densityall}
n_{\sigma}\!=&
\langle \psi_{0}|\hat a^{\dagger}_{\sigma}(z)\hat a_{\sigma}(z)|\psi_{0}\rangle
\\ \nonumber
\!=&
\frac{
|\alpha|^2|e^{i\phi/2}U_{\sigma \mu}
\!+\!
e^{-i\phi/2}U_{\sigma \nu}|^2
+2|\beta|^2(|U_{\sigma \mu}|^2
\!+\!
|U_{\sigma \nu}|^2)
}
{{\cal N}^2}
\end{align}
is plotted in~\rf{fig:symmetric}. The 1-photon sector interference pattern is phase $\phi$-dependent.

\begin{figure}[h]
\includegraphics[width=0.99\columnwidth]{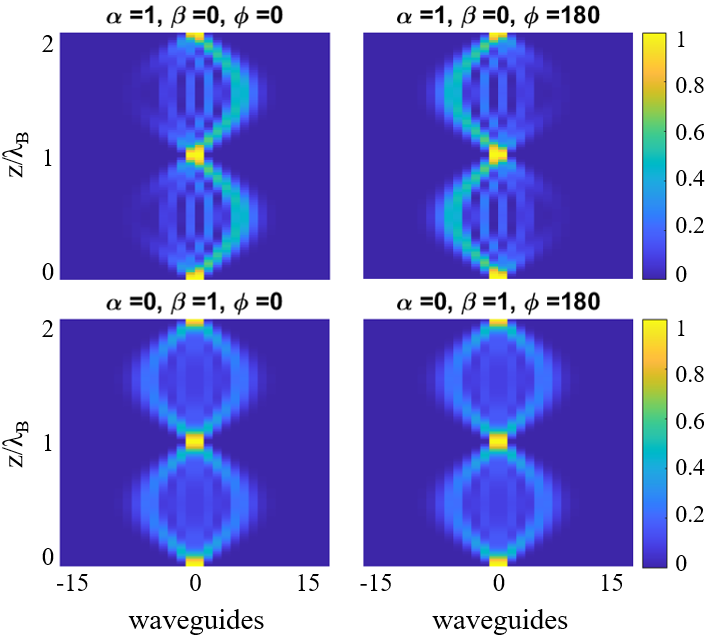}
\caption{
Photon density evolution for input N00N states - light propagates in the $z$ direction, tunneling between the 30 waveguides spanning the platform, using $B=1$, $C=1.75$ as an example.}
\label{fig:symmetric}
\end{figure}

Varying the phase, one controls which side of the platform the bright fringe resides for the $N=1$ N00N state input. For the $N=2$ N00N input, the resulting interference exhibits photon density peaks on both sides, irrespective of phase (note  this changes if one includes $|11\rangle$ states in the input, see~\ra{11interference}). In these plots the N00N states are input into the center $\mu,\nu=0,1$ waveguides, and the propagation direction scales in units of Bloch oscillation period.

\section{Bloch oscillations in asymmetric loss platforms}
\label{asymmetric}

\subsection{Setup and theory}

We modify the above Bloch oscillation platform to include a zone of total loss in one half of the waveguides as shown in the illustrative example in~\rf{fig:schematicWidths}. This cut is placed at $\lambda_B/2$, the half Bloch period at which the fringes are farthest apart from the central input/output waveguides. To realize such a cut, we envision e.g. a reflective Bragg scattering lattice, or an adiabatic siphoning of the waveguides out of  the Bloch oscillation array. The range in $z$ (propagation direction) required to make the cut should be well below the Bloch period so as not to impede the oscillations near $z=0,\lambda_B$ where the photon density converges to the central waveguides. Note that while~\rf{fig:schematicWidths} illustrates waveguides of varying width to impart $B$ (changing effective index), one may also consider the approach in~\cite{Lebugle_2015} using concentric arc waveguides.

\begin{figure}
\includegraphics[width=0.99\columnwidth]{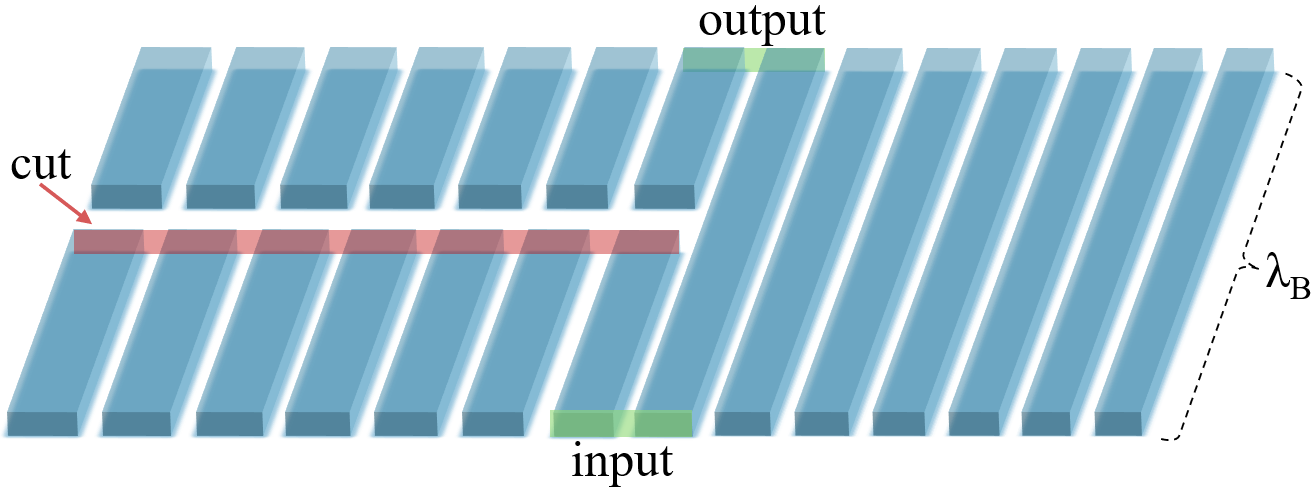}
\caption{Example of an asymmetric loss platform (not to scale). Two waveguides serve as inputs and outputs (green), and the cut (red) labels 0 transmission. New waveguides are added after the cuts to minimize edge effects.}
\label{fig:schematicWidths}
\end{figure}

To simulate photon densities with this cut, we compute the evolution of the reduced density matrix. We first apply the spatial evolution to the initial state~\req{psiInitial}: $|\psi(z)\rangle={\mathrm e}^{-i \hat Hz }|\psi_0\rangle$. To obtain the spatial evolution, we use the Green's function solution:
\begin{equation}\label{evolution}
    \begin{split}
        |\psi(z)\rangle&={\mathrm e}^{-i \hat Hz }|\psi_0\rangle\\
        &=\frac{1}{\cal N}{\mathrm e}^{-i \hat Hz }\bigg(\alpha \left(\hat{a}^{\dagger}_{\mu}(0)+{\mathrm e}^{2 i \phi}\hat{a}^{\dagger}_{\nu}(0)\right)+\\
        &\frac{\beta}{\sqrt{2}}\left(\hat{a}^{\dagger}_{\mu}(0)^2+{\mathrm e}^{2 i \phi}\hat{a}^{\dagger}_{\nu}(0)^2\right)\bigg){\mathrm e}^{i \hat Hz }{\mathrm e}^{-i \hat Hz }|0^{\otimes N}\rangle\\
        &=\frac{1}{\cal N}\bigg(\alpha \left(\hat{a}^{\dagger}_{\mu}(-z)+{\mathrm e}^{2 i \phi}\hat{a}^{\dagger}_{\nu}(-z)\right)+\\
        &\frac{\beta}{\sqrt{2}}\left(\hat{a}^{\dagger}_{\mu}(-z)^2+{\mathrm e}^{2 i \phi}\hat{a}^{\dagger}_{\nu}(-z)^2\right)\bigg)|0^{\otimes N}\rangle\\
    \end{split}
\end{equation}
where
\begin{align}
\hat a^{\dagger}(-z)={\mathrm e}^{-i\hat Hz }\hat a^{\dagger}(0){\mathrm e}^{i \hat Hz }
\end{align}
in the Heisenberg picture, and $\hat a^{\dagger}(-z)$ is given by~\req{aEvolve}.

To simulate the cut in~\rf{fig:schematicWidths}, we  evolve the wavefunction~\req{evolution} to the cut position 
\begin{align}
z_c=\lambda_B/2
\;.
\end{align}
The density matrix at this point is given by $\rho(z=z_c)=|\psi(z_c)\rangle\langle\psi(z_c)|$. Total loss on half the waveguides is implemented by tracing over the modes in those waveguides. We therefore compute the reduced density matrix 
\begin{align}
\rho^{\rm cut}_{i\geq1}(z=z_c)= 
\sum_{n_{i<1}}\langle n_{-N/2}...n_0|\psi\rangle\langle \psi|n_{-N/2}...n_0\rangle
\;.
\end{align}

To obtain a time evolving state after the cut, it is most convenient to reintroduce the cut waveguides as vacuum modes. We can then use the same equations of motion to  evolve the state, while avoiding edge effects. This reintroduction amounts to the tensor product
\begin{align}
\label{reintroduceGuides}
\rho(z_c+\epsilon)= 
|0_{-N/2}...0_0\rangle\langle0_{-N/2}...0_0|\otimes\rho^{\rm cut}_{i\geq 1}(z=z_c)
\;,
\end{align}
where infinitesimal $\epsilon\ll\lambda_B$. An experiment will certainly involve a finite distance in place of $\epsilon$, in which case the Bloch period needs to be made sufficiently long in comparison for these simulations to apply.

We can now evolve the density matrix past the cut region:
\begin{align}
\label{densityEvolution}
\rho(z>z_c)= 
{\mathrm e}^{i H(z-z_c) }\rho(z_c+\epsilon){\mathrm e}^{-i H(z-z_c) }
\;.
\end{align}
The evolution is carried out up to the full Bloch period 
\begin{align}
z_{\rm out}=\lambda_B
\;,
\end{align}
which we recall is the point where the zero loss Bloch oscillations in~\rf{fig:symmetric} refocus into the initial state distribution. This full calculation can be done analytically (\ra{N1check} and \ra{AnalN00N}); however, for higher $N>1$ N00N states, numerical simulation is more efficient. Although the initial state is not fully recovered here, $z_{\rm out}=\lambda_B$ remains the optimal point for filtration.

Extracting the output at the two central waveguides requires the following reduced density matrix, given by the partial trace
\begin{align}
\label{rhoOut}
\rho^{\rm out}=&\;
\sum_{n_i}
\langle n_{-N/2}...n_{-1}n_{2}...n_{N/2}|
\\ \nonumber 
&\;\quad \times 
\rho(z_{\rm out})|n_{-N/2}...n_{-1}n_{2}...n_{N/2}\rangle
\;.
\end{align}
From the diagonal elements we extract the conditional probabilities for the 1-photon and 2-photon subspaces:
\begin{align}
\label{P1P2}
P_1=&\;
\frac{\langle10|\rho^{\rm out}|10\rangle+\langle01|\rho^{\rm out}|01\rangle}{1-P_0}
\;,
\\ \nonumber
P_2=&\;
\frac{\langle20|\rho^{\rm out}|20\rangle +\langle02|\rho^{\rm out}|02\rangle +\langle11|\rho^{\rm out}|11\rangle}{1-P_0}
\;,
\end{align}
where the probability of having vacuum $P_0=\langle00|\rho^{\rm out}|00\rangle$. $P_1$ and $P_2$ determine the ratio of 1-photon to 2-photon probabilities, which we control via the input phase $\phi$.

\subsection{Simulation}

We compute the conditional probabilities in~\req{P1P2} by simulating the density matrix evolution in MATLAB. We optimize the degree of filtration by varying waveguide parameters $B$ (controlling the Bloch period) and $C$ (evanescent coupling strength). In particular we vary the ratio $C/B\propto \lambda_B/L_t$, which determines the spatial extent of the bright fringes at the half Bloch period: the larger the ratio, the larger the transverse spread.

We simulate a 6-waveguide platform, with total loss in 3 waveguides at the half Bloch period. The cut is placed on the left side of the platform where, for positive $B$, the linearly growing index profile means smaller effective indices. Since the $N=1$ sector fringes are localized here when $\phi=\pi$ (see~\rf{fig:symmetric}), this is the optimal phase for $N=2$ subspace filtration. For numerical efficiency, we truncate the maximum photon number in a given mode to be two. In~\rf{fig:varyz} we plot the conditional probabilities with and without cuts compared as functions of propagation distance $z$. Using  $C/B=0.3$ as an example, we find $>80\%$ filtering of the 2-photon sector ($|20\rangle$, $|02\rangle$ and $|11\rangle$) in the two output waveguides, suppressing the 1-photon sector ($|10\rangle$ and $|01\rangle$).

\begin{figure}[h]
\includegraphics[width=0.99\columnwidth]{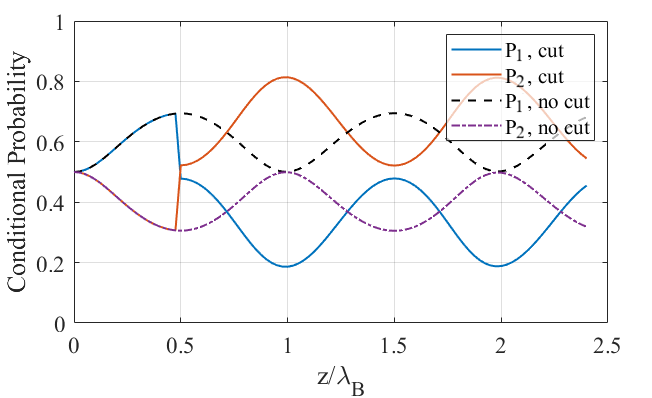}
\caption{Conditional probabilities (\req{P1P2}) with and without cuts at the half Bloch period ($z=\lambda_B/2$), plotted as functions of normalized position. The input $N=1$ and $N=2$ N00N states are applied with equal probability ($\alpha=\beta=1$, see~\req{psiInitial}), with $C/B=0.3$ and phase $\phi=\pi$ to achieve $>80\%$ filtration of the $N=2$ sector.}
\label{fig:varyz}
\end{figure}

To put this into experimental context, we consider parameters compatible with a platform of SiN waveguides in SiO$_2$ cladding. One may envision a design tailoring the cladding gap widths to fix the distance $L_t=200\mu{\rm m}$ for near-total power transfer between waveguides, giving $C=\pi/2L_t=0.008\mu{\rm m}^{-1}$. We then vary $B$ via the waveguide widths, or by modifying the radii of curvature in the case of concentric arc geometries. For the ratio $C/B=0.3$ we then have the step in propagation constant $B=0.027\mu{\rm m}^{-1}$. Considering e.g. a wavelength of $\lambda=1550$nm and an  effective index $n_{\rm eff}=1.5$, this amounts to $\Delta n_{\rm eff}\sim0.007$. Recalling $\Delta n_{\rm eff}=n_{\rm eff}d/R$~\cite{Lebugle_2015} for the concentric arc case, these parameters are feasible for waveguide spacings $d\sim{\mathcal O}(\mu{\rm m})$ and radii $R\sim{\mathcal O}({\rm mm})$.

This result can also be adapted to larger waveguides or fiber cores. We recall from the Green's function in~\req{Ugreenform} that $C/B$ governs the photon density transverse spread, while $B$ alone determines the Bloch period. Indeed, it is possible to do normalization of $z$ so that only $C/B$ dependence remains in the equations of motion. Thus~\rf{fig:varyz} holds for different $B$ and $C$ so long as their ratio is constant.

It is also possible to filter the 1-photon sector, tuning the phase to switch between 1 and 2-photon dominated cases. In~\rf{fig:varyphi} we plot $P_1$ and $P_2$ at the full Bloch period while varying the phase. At $\phi=\pi$ we have the maximum 2-photon filtration seen in~\rf{fig:varyz}, while at $\phi\sim\pi/2,3\pi/2$ we have the maximum 1-photon sector filtering.

\begin{figure}
\includegraphics[width=0.99\columnwidth]{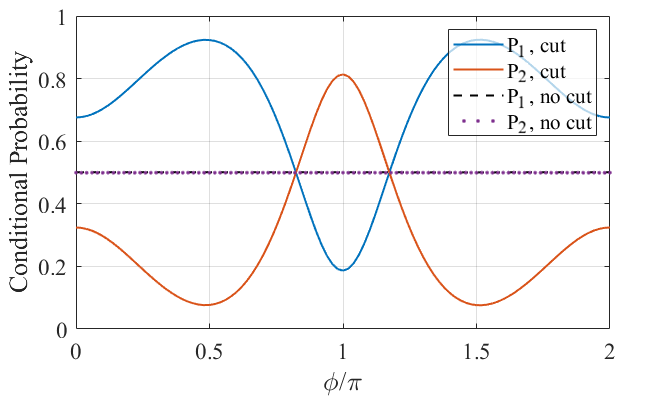}
\caption{
Plot of conditional probabilities with and without cuts as a function of input phase. The same parameters as in~\rf{fig:varyz} apply: $\alpha=\beta=1$ and $C/B=0.3$. The probabilities are computed at the full Bloch period $z=\lambda_B$. 
}
\label{fig:varyphi}
\end{figure}

\begin{figure}
\includegraphics[width=0.99\columnwidth]{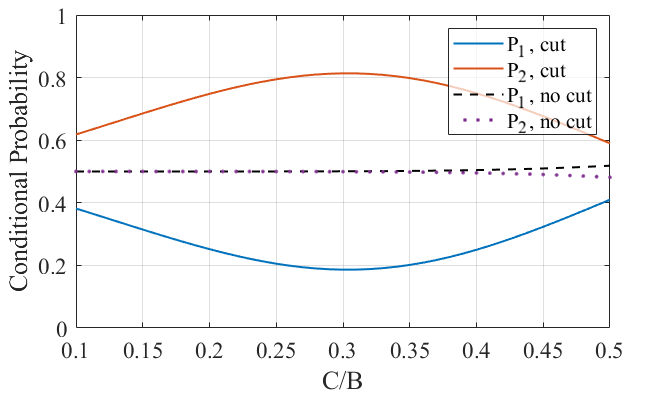}
\caption{
Plot of conditional probabilities with and without cuts as a function $C/B$, with fixed $\alpha=\beta=1$, $z=\lambda_B$ and $\phi=\pi$. 
}
\label{fig:varyCB}
\end{figure}

We also consider different $C/B$ ratios for 2-photon sector filtration at the optimal phase and $z$ positions. In~\rf{fig:varyCB} we plot $P_1$ and $P_2$ at fixed $z=\lambda_B$ and $\phi=\pi$, finding the optimal parameters at $C/B\sim0.3$. Note that this optimization only applies to the 2-photon sector - for 1-photon filtration the optimal ratio is larger. However, edge effects become important starting at $C/B\sim0.5$, where in~\rf{fig:varyCB} the uncut results at $z=\lambda_B$ deviate from the equal probability inputs. This requires simulation of larger waveguide arrays, which we reserve for follow-up work.

Furthermore, we vary input parameters $\alpha$ and $\beta$ to test different $N=1$ N00N state dominated inputs. We consider ratios $\alpha/\beta$ such that the initial $N=1$ N00N states comprise $50\%,90\%,99\%$ and $99.9\%$ of the input, with results shown in~\rf{fig:varyIP}. The metric used here is the mode participation ratio $(P_2(z)/P_1(z))/(P_2(0)/P_1(0))$. We find that this filtration procedure is more effective for $N=1$ N00N state dominated inputs. This relative amplification of the $P_2$ value is accomplished by a relative loss of total intensity. In the case where the initial state has a population of $90\%$ N=1 N00N states, we calculate the loss to be $94.1\%$ roughly corresponding to near total loss of $P_1$ states and a loss of half the initial $P_2$ states, as would be expected based on the symmetrical behavior shown in \rf{fig:symmetric}. 

\begin{figure}
\includegraphics[width=0.97\columnwidth]{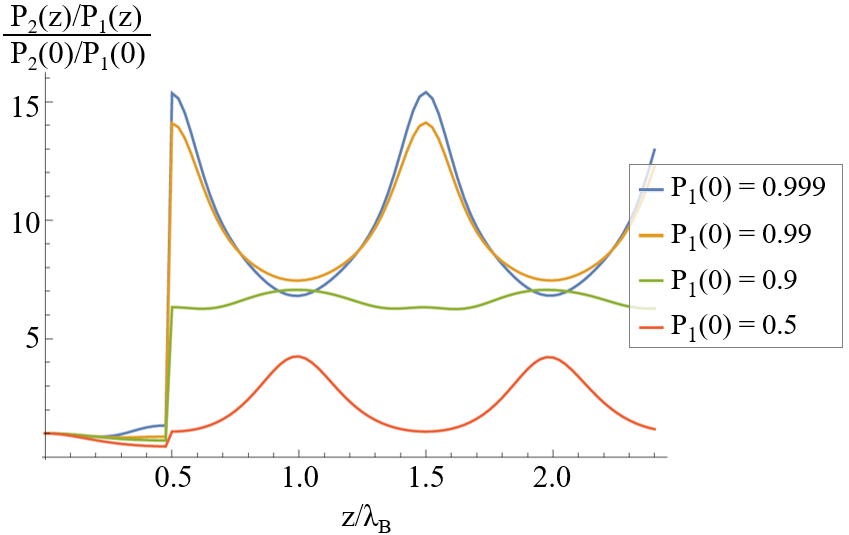}
\caption{
Plot of normalized mode participation ratios as a function of $z$, with the initial fraction of the population in the $N=1$ N00N state being varied, while $C/B=0.3$ and $\phi=\pi$ for $N=2$ filtration. Normalized efficiencies for these different populations of photon number states have been calculated and are found in \ra{analysisEfficiency}.
}
\label{fig:varyIP}
\end{figure}

\section{Conclusion}
\label{concl}

In conclusion, we have computed the evolution of varying populations of $N=1$ and $N=2$ N00N states in an array of waveguides with linearly growing propagation constants and asymmetric losses. Our approach leverages the asymmetrical phase dependent fringe distribution of $N=1$ N00N state Bloch oscillations. Filtering them out via siphoned waveguides, we obtain output states with higher proportions of the $N=2$ or $N=1$ photon subspaces when varying the N00N state phase and conditioning out the vacuum. While the $N=2$ transmission for 2-photon subspace dominated configurations is currently $\sim40\%$, see~\ra{analysisEfficiency}, this approach is fundamentally different from destructive detection and may provide a new avenue for non-destructive routing.

This theoretical filtration approach relies on preparation of the initial state~\req{psiInitial} with high fidelity and low loss, a separate experimental challenge. The performance of this filter for mixed initial states remains unexplored, but it is likely that filtration efficiency will be degraded as phase information, i.e. the asymmetry between $N=1$ and $N=2$ Bloch oscillations, is lost in the mixing. However, for 1-photon sector dominated initial states (see~\rf{fig:varyIP}) we expect significant $N=2$ sector filtration to be achievable even with large loss.

Our approach opens the path to high-fidelity dynamic switching between photon numbers via phase tuning. Further improvement to the filtration approach may be explored by varying the number of lossy waveguides, or extracting the signal from non-central waveguides. We may also consider partial loss e.g. with photonic crystals, instead of the total loss considered thus far via siphoning, where the interference effects of reflected light will also need to be simulated. Beyond the uniform half-platform cut simulated here, one may envision staggered or other optimized waveguide number dependent loss arrangements.

\bibliographystyle{apsrev4-2}
\bibliography{refsfilter}

\appendix

\section{2-photon sector interference}
\label{11interference}

We revisit the Bloch oscillations for the zero loss case in~\rs{overview}. Including the $|11\rangle$ state in~\req{psiInitial},
\begin{align}
\label{psiinitial11}
|\psi_0\rangle=&\;
\frac{1}{\cal N}
\Big\{
\alpha\Big(|1_\mu0_{\nu}\rangle+e^{i\phi}|0_{\mu}1_{\nu}\rangle\Big)
\\ \nonumber
&\;
+\beta\Big(|2_{\mu}0_{\nu}\rangle+e^{2i\phi}|0_{\mu}2_{\nu}\rangle\Big)
+\sqrt2 \gamma e^{i\phi}|11\rangle
\Big\}
\;, 
\end{align}
where now ${\cal N}=\sqrt{2(|\alpha|^2+|\beta|^2+|\gamma|^2)}$. The photon density 
\begin{align}
\label{densityall11}
n_{\sigma}=&\;
\frac{1}{{\cal N}^2}
\Big\{
|\alpha|^2|e^{i\phi/2}U_{\sigma \mu}+e^{-i\phi/2}U_{\sigma \nu}|^2
\\ \nonumber
&\;\quad
+2\Big(|\beta|^2+|\gamma|^2\Big)\Big(|U_{\sigma \mu}|^2+|U_{\sigma \nu}|^2\Big)
\\ \nonumber
&\;\quad
+2
\Big(\gamma^*\beta+\gamma\beta^*\Big)
\Big(U_{\sigma \mu}U^\dagger_{\sigma \nu} e^{i\phi}+ U_{\sigma \nu}U^\dagger_{\sigma \mu} e^{-i\phi}\Big)
\Big\}
\;,
\end{align}
where the last term accounts for the additional interference between $|11\rangle$ and $N=2$ N00N states.

Considering e.g. low intensity coherent state inputs (see~\cite{Kim_2021}) where we may neglect higher than 2 photon number states, we envision a $|20\rangle$ state applied to a beam-splitter ($|2\rangle$ in one port and vacuum in the other). The result is a superposition of $N=2$ N00N and $|11\rangle$ states, with a fixed $|\beta|=|\gamma|$. Applying the beam-splitting output to the input waveguides $\mu,\nu$ of the Bloch oscillation platform, we find the 2-photon sector intensity distribution peaks on the same side as that of the 1-photon sector seen in~\rf{fig:symmetric}. The inability to distinguish between $N=1$ and $N=2$ sectors is expected here, since a non-Hermitian linear optics setup may only filter photon numbers from a pre-prepared quantum state, and not from a classical state (or in this case, Fock state components of a coherent state).

\section{Analytical example of $N=1$ N00N state density matrix in a no loss platform}
\label{N1check}

As a simpler example we consider the $N=1$ N00N state as an initial state. The state evolution is given by:
\begin{equation}
    |\psi(z)\rangle=\frac{1}{\sqrt{2}}\left(\delta_{\mu \alpha}+\delta_{\nu \alpha}\right)U_{\alpha \mu'}(-z)a^{\dagger}_{\mu'}(0)|0^{\otimes N}\rangle.
\end{equation}
Then we have:
\begin{align}
&
    \langle n_1\dots n_{\gamma-1}n_{\gamma+1} \dots n_N|\psi(z)\rangle
    \\ \nonumber
    =&\;\frac{1}{\sqrt{2}}\left(\delta_{\mu \alpha}+\delta_{\nu \alpha}\right)
    \\ \nonumber &\; \times
    \Big\{
    U_{\alpha \mu'}(-z)\delta_{n_1}\dots \delta_{n_{\mu'}1}\dots \delta_{n_N 0}\left(1-\delta_{\mu'\gamma}\right)|0\rangle_{\gamma}
    \\ \nonumber &\qquad +
    U_{\alpha \gamma}(-z)\delta_{n_1 0}\dots \delta_{n_N 0}|1\rangle_{\gamma}
    \Big\}
    \;.
\end{align}
We can use these to calculate $\langle n_\gamma \rangle=Tr[n_{\gamma}\rho_{\gamma}]$. 
We now calculate the matrix elements of $\rho_{\gamma}$:
\begin{align}
    \rho^{11}_{\gamma}=&\;
    \frac{1}{2}\left|(\delta_{\mu \alpha}+\delta_{\nu \alpha})U_{\alpha \gamma}(-z)\right|^2
    \\ \nonumber
    \rho^{01}_\gamma=&\;0
\\ \nonumber 
    \rho^{00}_{\gamma}=&\;
    1-\rho^{11}_{\gamma}
    \;.
\end{align}

As a consistency check, we compute $\langle a^{\dagger}_\mu a_\mu\rangle$. Given this density matrix, we have:
\begin{equation}
\begin{split}
    \langle a^{\dagger}_\mu a_\mu\rangle
    &={\mathrm{Tr}}(\rho_{\mu}a^{\dagger}_\mu a_\mu) =\rho^{11}_{\gamma}
    \\&=\frac{1}{2}\left|(\delta_{\mu \alpha}+\delta_{\nu \alpha})U_{\alpha \gamma}(-z)\right|^2\\
    &=\frac{1}{2}\left|U_{\mu \gamma}+U_{\nu \gamma}\right|^2
    \end{split}
    \;,
\end{equation}
recovering the result of Bromberg et al.~\cite{Bromberg_2010}.
Summing the expected photon counts over all waveguides i.e. $\sum_{\gamma}{\mathrm{Tr}\left[\hat{n}_{\gamma}\rho_{\gamma}\right]}$, it is straightforward to verify that the total photon number quantity is conserved.

\section{Analytical form of $N=1$ N00N state evolution with cuts}\label{AnalN00N}

We consider an analytical solution to $N=1$ N00N state Bloch oscillations in a cut platform. The state vector directly before the cut is given by:

\begin{equation}
    |\psi(z)\rangle=\frac{1}{\sqrt{2}}\left(\delta_{\mu \alpha}+\delta_{\nu \alpha}\right)U_{\alpha \mu'}(-z)a^{\dagger}_{\mu'}|0^{\otimes N}\rangle
    \;. 
\end{equation}
Again, we are using Einstein summation notation, $\mu$ and $\nu$ are the input waveguides for the N00N state, and $N$ is the number of waveguides in the register (assumed to be large enough that edge effects can be ignored). The system evolves under the Hamiltonian evolution given by \ref{eqn:HeisenbergEquations}. Now, at $z=z_c$, we cut waveguides on the left side of the platform ($\mu \leq 0$). This is accomplished by tracing over these degrees of freedom. We have:
\begin{align}
&
\langle n_{-N/2}...n_0|\psi\rangle
\\ \nonumber 
&=
\frac{1}{\sqrt{2}}\left(\delta_{\mu \alpha}+\delta_{\nu \alpha}\right)U_{\alpha \mu'}(-z_c)\langle n_{-N/2}...n_0|a^{\dagger}_{\mu'}|0^{\otimes N}\rangle
\\ \nonumber
&=
\frac{1}{\sqrt{2}}\left(\delta_{\mu \alpha}+\delta_{\nu \alpha}\right)U_{\alpha \mu'}(-z_c)\delta_{n_{-N/2}0}...\delta_{n_{\mu'} 1}...\delta_{n_0 0}
\\ \nonumber
&\qquad
\times
\left(1-\Theta(\mu')\right)|0_1...0_{N/2}\rangle
\\ \nonumber 
&\quad +
\frac{1}{\sqrt{2}}\left(\delta_{\mu \alpha}+\delta_{\nu \alpha}\right)U_{\alpha \mu'}(-z_c)\delta_{n_{-N/2}0}...\delta_{n_0 0}
\\ \nonumber
&\qquad
\times
\Theta(\mu')|0_1...1_{\mu'}...0_{N/2}\rangle
\;.
\end{align}
$\Theta$ is the Heaviside function, defined such that $\Theta(0)=0$. We now compute $\sum_{n_{i<1}}\langle n_{-N/2}...n_0|\psi\rangle\langle \psi|n_{-N/2}...n_0\rangle$ to obtain
\begin{align}
&
\rho_{i\geq 1}(z=z_c)
\\ \nonumber
&=
\frac{1}{2}\sum_{\mu'}|U_{\mu \mu'}(-z_c)+U_{\nu \mu'}(-z_c)|^2\left(1-\Theta(\mu')\right)
\\ \nonumber
&\qquad \times
|0_1...0_{N/2}\rangle\langle 0_1...0_{N/2}|
\\ \nonumber
&\quad+
\frac{1}{2}\sum_{\mu' \mu''}\left(U_{\mu \mu'}(-z_c)+U_{\nu \mu'}(-z_c)\right)
\\ \nonumber
&\qquad \times
\left(U^*_{\mu \mu''}(-z_c)+U^*_{\nu \mu''}(-z_c)\right) \Theta(\mu')\Theta(\mu'')
\\ \nonumber
&\qquad \times
|0_1...1_{\mu'}...0_{N/2}\rangle\langle 0_1...1_{\mu''}...0_{N/2}|
\;.
\end{align}
Following~\req{reintroduceGuides} we reintroduce the waveguides after the cut:
\begin{align}
&
\rho_{i\geq 1}(z=z_c)
\\ \nonumber 
&=
\frac{1}{2}\sum_{\mu'}|U_{\mu \mu'}(-z_c)+U_{\nu \mu'}(-z_c)|^2\left(1-\Theta(\mu')\right)
\\ \nonumber
&\qquad \times
|0_{-N/2}...0_{N/2}\rangle\langle 0_{-N/2}...0_{N/2}|
\\ \nonumber
&\quad +
\frac{1}{2}\sum_{\mu' \mu''}\left(U_{\mu \mu'}(-z_c)+U_{\nu \mu'}(-z_c)\right)
\\ \nonumber
&\qquad \times
\left(U^*_{\mu \mu''}(-z_c)+U^*_{\nu \mu''}(-z_c)\right) \Theta(\mu')\Theta(\mu'')
\\ \nonumber
&\qquad \times
|0_{-N/2}...1_{\mu'}...0_{N/2}\rangle\langle 0_{-N/2}...1_{\mu''}...0_{N/2}|
\;.
\end{align}

We now use this density matrix as an initial condition for further evolution after the cut. Expressing the state in terms of creation and annihilation operators and using the solution~\req{aEvolve} from~\cite{Bromberg_2010}, the result is:
\begin{align}
& \rho_{i\geq 1}(z>z_c)
\\ \nonumber
&=\frac{1}{2}\sum_{\mu'}|U_{\mu \mu'}(-z_c)+U_{\nu \mu'}(-z_c)|^2\left(1-\Theta(\mu')\right)
\\ \nonumber
&\qquad \times
|0_{-N/2}...0_{N/2}\rangle\langle 0_{-N/2}...0_{N/2}|
\\ \nonumber
&+
\frac{1}{2}\sum_{\mu' \mu''}\left(U_{\mu \mu'}(-z_c)+U_{\nu \mu'}(-z_c)\right)
\\ \nonumber
&\qquad \times\left(U^*_{\mu \mu''}(-z_c)+U^*_{\nu \mu''}(-z_c)\right)
\\ \nonumber
&\qquad \times
\Theta(\mu')\Theta(\mu'')U_{\mu' \delta}(-(z-z_c))a^{\dagger}_{\delta}
\\ \nonumber
&\qquad \times
|0_{-N/2}...0_{N/2}\rangle\langle 0_{-N/2}...0_{N/2}|U^*_{\mu'' \epsilon}(-(z-z_c))a_{\epsilon}
\;.
\end{align}

The density matrix as a function of z is therefore a piecewise function:
\begin{equation}
    \rho(z)=
    \begin{cases}
        |\psi(z)\rangle\langle\psi(z)|& z<z_c\\
        \rho_{i\geq 1}(z)& z>z_c
    \end{cases}
    \;.
\end{equation}
After the cut, the reduced density matrix for a waveguide ($\gamma$) is given by:
\begin{align}
    \rho^{(0,0)}_{\gamma}=&\;
    \frac{1}{2}\sum_{\mu'}|U_{\mu \mu'}(-z_c)+U_{\nu \mu'}(-z_c)|^2\left(1-\Theta(\mu')\right)
    \nonumber \\ &\; +
    \frac{1}{2}\sum_{\mu' \mu''}\left(U_{\mu \mu'}(-z_c)+U_{\nu \mu'}(-z_c)\right)
    \\ \nonumber &\; \times
    \left(U^*_{\mu \mu''}(-z_c)+U^*_{\nu \mu''}(-z_c)\right)\Theta(\mu')\Theta(\mu'')
    \\ \nonumber &\; \times
    U_{\mu' \delta}(z_c-z)\left(1-\delta_{\delta \gamma}\right)\left(1-\delta_{\epsilon \gamma}\right)U^*_{\mu'' \epsilon}(z_c-z)
    \;,
\end{align}
\begin{align}
    \rho^{(0,1)}_{\gamma}=&\;
    \frac{1}{2}\sum_{\mu' \mu''}\left(U_{\mu \mu'}(-z_c)+U_{\nu \mu'}(-z_c)\right)
    \\ \nonumber &\; \times
    \left(U^*_{\mu \mu''}(-z_c)+U^*_{\nu \mu''}(-z_c)\right)\Theta(\mu')\Theta(\mu'')
    \\ \nonumber &\;\times
    U_{\mu' \delta}(-(z-z_c))\left(1-\delta_{\delta \gamma}\right)U^*_{\mu'' \gamma}(-(z-z_c))
    \;,
\end{align}
\begin{align}
    \rho^{(1,1)}_{\gamma}=&\;
    \frac{1}{2}\sum_{\mu' \mu''}\left(U_{\mu \mu'}(-z_c)+U_{\nu \mu'}(-z_c)\right)
    \\ \nonumber &\; \times
    \left(U^*_{\mu \mu''}(-z_c)+U^*_{\nu \mu''}(-z_c)\right)\Theta(\mu')\Theta(\mu'')
    \\ \nonumber &\; \times
    U_{\mu' \gamma}(-(z-z_c))U^*_{\mu'' \gamma}(-(z-z_c))
    \;.
\end{align}
Analysis of the evolution of $N>1$ photon sector initial states is also possible, but at this point simulation is more efficient.

\section{Numerical Analysis of Two Photon Efficiency}
\label{analysisEfficiency}

Here we numerically analyze the probability of the state being in various number states after evolving in the waveguide array with cuts. Using the output reduced density matrix~\req{rhoOut}, we define: $E_1=\langle01|\rho^{\rm out}|01\rangle +\langle10|\rho^{\rm out}|10\rangle$, and $E_2=\langle02|\rho^{\rm out}|02\rangle +\langle20|\rho^{\rm out}|20\rangle+ \langle11|\rho^{\rm out} |11\rangle$. These quantities have been computed for a range of $z/\lambda_B$ using the same waveguide configurations as in~\rf{fig:varyz}, with phase $\phi=\pi$ chosen to suppress the 1-photon sector, and with different proportions of initial $N=1$ N00N states. Results are shown in \rf{fig:varyIPAbs}. We normalize $E_2$ and $E_1$ by their initial values. Even with relatively high proportions of initial $N=1$ N00N states, $E_2$ exhibits a high efficiency due to the $N>1$ N00N states evolving symmetrically across the array. On the other hand, $E_1$ experiences more significant suppression as the initial population of $N=1$ N00N states becomes larger, because these states evolve asymmetrically in the waveguide array.

\begin{figure}
\includegraphics[width=0.97\columnwidth]{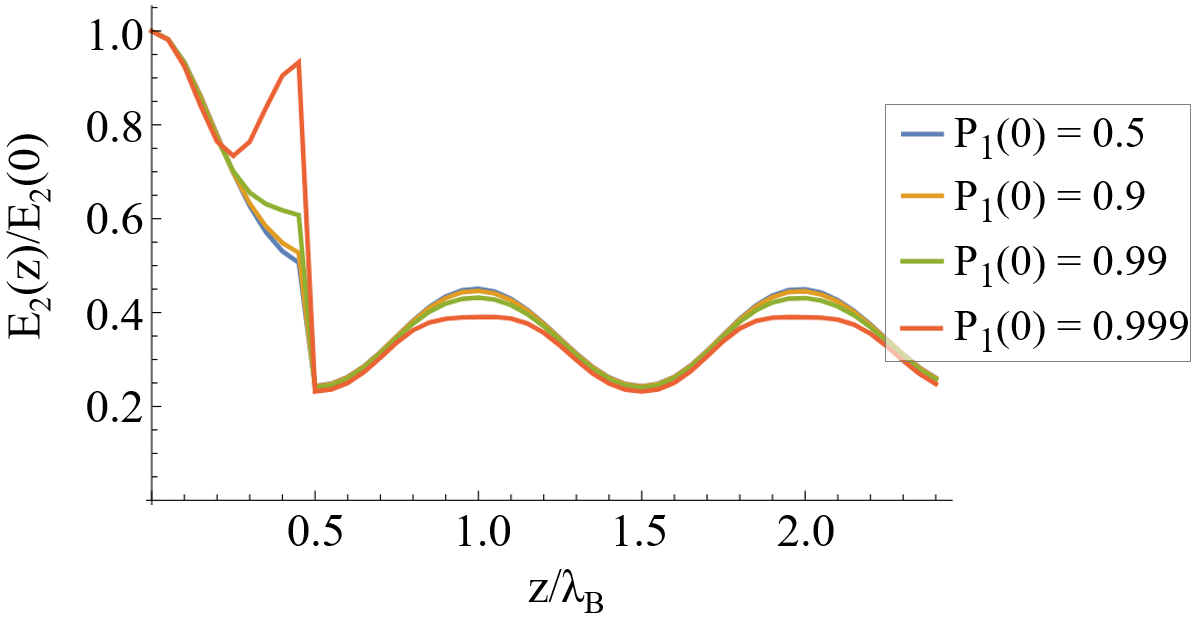}
\includegraphics[width=0.83\columnwidth]{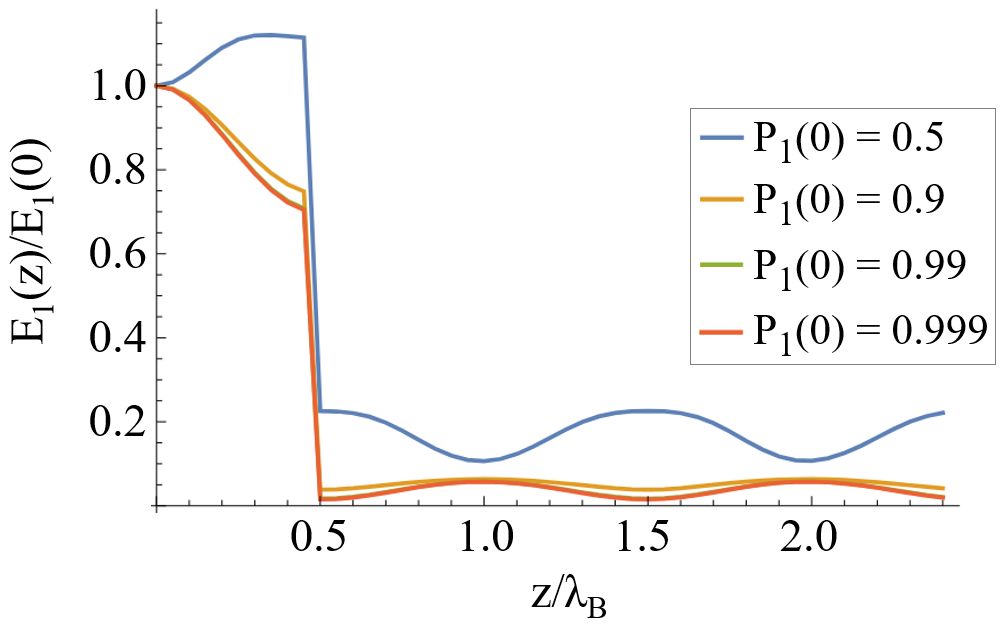}
\caption{
Top: normalized efficiency ratio $E_2/E_2(0)$ and bottom: normalized efficiency ratio: $E_1/E_1(0)$. Reductions are seen close to the location of the cut. 
}
\label{fig:varyIPAbs}
\end{figure}

\end{document}